\documentclass[journal,a4paper,english]{IEEEtran}

\usepackage[nofonts,nobabel]{mh_basic}

\usepackage[cmex10]{amsmath}
\usepackage{mh_math,mh_tikz}
\usetikzlibrary{calc,decorations,decorations.pathreplacing}
\usepackage{pgfplots}
\usepackage{breakurl}

\usepackage{algorithm} 
\usepackage{algpseudocode}
\newtheorem{theorem}{Theorem}
\newtheorem{lemma}{Lemma}
\newtheorem{observation}{Observation}

\newcommand{\etal}{{\itshape et al.}\xspace}
\newcommand{\defn}[1]{\textit{#1}}

\usepackage[labelformat=simple]{subcaption}

\newcommand{\calP}{\mathcal P}
\newcommand{\calQ}{\mathcal Q}
\newcommand{\Ppath}{\calP_{\text{path}}}

\mathtoolsset{showonlyrefs=true}

\begin{document}
\sloppy
\title{Towards An Exact Combinatorial Algorithm for LP Decoding of Turbo Codes}

\author{
  \IEEEauthorblockN{Michael Helmling and Stefan Ruzika}
  \IEEEauthorblockA{
    Mathematical Institute\\
    University of Koblenz-Landau, Campus Koblenz\\
    56070 Koblenz, Germany\\
    Email: \{helmling, ruzika\}@uni-koblenz.de
    }%
	\thanks{
		The authors are with the Optimization Research Group, Department of Mathematics,
		University of Kaiserslautern, Erwin-Schroedinger-Strasse, 67663 Kaiserslautern, Germany.
		Email: \{helmling, ruzika\}@mathematik.uni-kl.de
	}%
	\thanks{
		We would like to acknowledge the Center for Mathematical and Computational Modelling (CM) and the DAAD
	}
}


\maketitle

\begin{abstract}
We present a novel algorithm that solves the turbo code LP decoding problem in a fininte number of steps by Euclidean distance minimizations, which in turn rely on repeated shortest path computations in the trellis graph representing the turbo code. Previous attempts to exploit the combinatorial graph structure only led to algorithms which are either of heuristic nature or do not guarantee finite convergence.
A numerical study shows that our algorithm clearly beats the running time, up to a factor of 100, of generic commercial LP solvers for medium-sized codes, especially for high SNR values.
\end{abstract}

\begin{IEEEkeywords}
LP decoding, turbo codes, combinatorial optimization
\end{IEEEkeywords}

%
\IEEEpeerreviewmaketitle

\section{Introduction}
Since its introduction by Feldman \etal in 2002 \cite{Feldman+02LPTurboDecoding}, Linear Programming based channel decoding has gained tremendous interest because of its analytical power---LP decoding exhibits the maximum likelihood (ML) certificate property \cite{Feldman03PhD}, and the decoding behavior is completely determined by the explicitly described “fundamental” polytope \cite{VontobelKoetter05GraphCover}---combined with noteworthy error-correcting performance and the availability of efficient decoding algorithms.

Turbo codes, invented by Berrou \etal in 1993 \cite{Berrou+93TurboCodes}, are a class of concatenated convolutional codes that, together with a heuristic iterative decoding algorithm, feature remarkable error-correcting performance.

While the first paper on LP decoding \cite{Feldman+02LPTurboDecoding} actually dealt with turbo codes, the majority of publications in the area of LP decoding now focus on LDPC codes \cite{Gallager62LDPC} which provide similar performance (cf.~\cite{Helmling+11MathProgDecoding} for a recent overview). Nevertheless, turbo codes have some analytical advantages, most importantly the inherent combinatorial structure by means of the trellis graph representations of the underlying convolutional encoders. ML Decoding of turbo codes is closely related to shortest path and minimum network flow problems, both being classical, well-studied topics in optimization theory for which plenty efficient solution methods exist. The hardness of ML decoding is caused by additional conditions on the path through the trellis graphs (they are termed \emph{agreeability constraints} in \cite{Feldman+02LPTurboDecoding}) posed by the turbo code's interleaver. Thus ML (LP) decoding is equivalent to solving a (LP-relaxed) shortest path problem with additional linear side constraints.

So far, two methods for solving the LP have been proposed: General purpose LP solvers like CPLEX \cite{CPLEX124} are based on the matrix representation of the LP problem. They utilize either the simplex method or interior point approaches \cite{Schrijver86LinearIntegerProg}, but do not exploit any structural properties of the specific problem. Lagrangian relaxation in conjunction with subgradient optimization \cite{Feldman+02LPTurboDecoding,Tanatmis+10Lagrangian}, on the other hand, utilizes this structure, but has practical limitations, most notably it usually converges very slowly.

This paper presents a new approach to solve the LP decoding problem exactly by an algorithm that exploits its graphical substructure, thus combining the analytical power of the LP approach with the running-time benefits of a combinatorial method which seems to be a necessary requirement for practical implementation. Our basic idea is to construct an alternative polytope in the space defined by the additional constraints (called \emph{constraints space}) and show how the LP solution corresponds to a specific point $z^\calQ_{\text{LP}}$ of that polytope. Then, we show how to computationally find $z^\calQ_{\text{LP}}$ by a geometric algorithm that relies on a sequence of shortest path computations in the trellis graphs.

The reinterpretation of constrained optimization problems in constraints space was first developed in the context of multicriteria optimization in \cite{Ruzika07PhD}, where it is applied to minimum spanning tree problems with a single side constraint. In 2010, Tanatmis \cite{Tanatmis10PhD} applied this theory to the turbo decoding problem. His algorithm showed a drastic speedup compared to a general purpose LP solver, however it only works for up to two constraints, while in real-world turbo codes the number of constraints equals the information length.

By adapting an algorithm by Wolfe \cite{Wolfe76NearestPoint} to compute in a polytope the point with minimum Euclidean norm, we are able to overcome these limitations and decode turbo codes with lengths of practical interest. The algorithm is, compared to previous methods, advantageous not only in terms of running time, but also gives valuable information that can help to improve the error-correcting performance. Furthermore, branch-and-bound methods for integer programming-based ML decoding depend upon fast lower bound computations, mostly given by LP relaxations, and can often be significantly improved by dedicated methods that evaluate combinatorial properties of the LP solutions. Since our LP decoder contains such information, it could also be considered a step towards IP-based algorithms with the potential of practical implementation.

\section{Background and Notation}
\subsection{Definition of Turbo Codes}
A $k$-dimensional subspace $C$ of the vector space $\F_2^n$ (where $\F_2=\{0,1\}$ denotes
the binary field), is called an $(n,k)$ binary linear block code, where $n$ is the block
length and $k$ the information (or input) length. One way to define a code is by an appropriate 
encoding function $e_C$, for which any bijective linear mapping from $\F_2^k$ onto $C$ qualifies.
This paper deals with turbo codes \cite{Berrou+93TurboCodes}, a special class
of block codes built by interconnecting (at least) two convolutional codes
(see \eg \cite{MacKay03InformationTheory}). For the sake of clear notation, we focus on turbo codes as used in the 3GPP LTE standard \cite{3GPP12LTE}—\ie, systematic,
parallely concatenated turbo codes with two identical terminated rate-$1$
constituent encoders—despite the fact that our approach is applicable to arbitrary turbo coding schemes. An in-depth covering of turbo code construction can be found in \cite{LinCostello04ECC}.

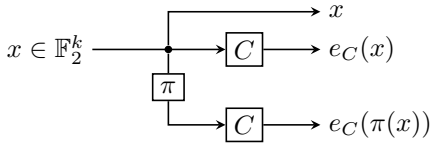
\begin{figure}\centering
    \begin{tikzpicture}[
	    	inner sep=1mm,
	    	sq/.style={rectangle,draw,semithick},
	    	arrow/.style={->,semithick},yscale=.5]
        \node[left] (u) at (0,0) {$x \in \F_2^k$};
        \node[fill,circle,inner sep=0mm,minimum size=1mm] (cross) at (1,0) {};
        \node[sq] (Ca) at (2,0) {$C$};
        \node[sq] (pi) at (1,-1) {$\pi$};
        \node[sq] (Cb) at (2,-2) {$C$};
        \draw[arrow] (u) -- (Ca);
        \draw[arrow] (pi) |- (Cb);
        \draw[semithick] (cross) -- (pi);
        \draw[arrow] (cross) |- ++(2,1) node[right] {$x$};
        \draw[arrow] (Ca) -- ++(1,0) node[right] {$e_C(x)$};
        \draw[arrow] (Cb) -- ++(1,0) node[right] {$e_C(\pi(x))$};
    \end{tikzpicture}
    \caption{Turbo encoder with two convolutional encoders $C_a$, $C_b$ and interleaver $\pi$.}\label{fig:turboSchematic}
\end{figure}

An $(n, k)$ turbo code $TC=TC(C,\pi)$ is defined by a \mbox{rate-$1$} convolutional $(n_C,k)$ code $C$ with constraint length $d$ and a permutation $\pi \in \mathbb{S}_k$ such that $n = k + 2\cdot n_C$.
Because we consider terminated convolutional codes only (\ie, there is a designated terminal state of the encoder), the final $d$ bits of the information sequence (also called the \defn{tail}) are not free to choose and thus can not carry any information.
Consequently, those bits together with the corresponding $d$ output bits are considered part of the output, which yields $n_C=k+2\cdot d$ and a code rate slightly below $1$.
Let $e_C: \F_2^k\longrightarrow \F_2^{n_C}$ be the associated encoding function. Then, the encoding function of $TC$ is defined as
\begin{equation}\begin{aligned}
	e_{TC}:\F_2^k &\longrightarrow \F_2^{k + 2\cdot n_C}\\
	e_{TC}(x) &=\left(x\;|\;e_C(x)\;|\;e_C(\pi(x))\right)
\end{aligned}\label{eq:e_TC}\end{equation}
where $\pi(x) = (x_{\pi(1)}, \dotsc, x_{\pi(k)})$.
In other words, the codeword for an input word $x$ is obtained by concatenating
\begin{itemize}
	\item a copy of $x$ itself,
	\item a copy of $x$ encoded by $C$, and
	\item a copy of $x$, permuted by $\pi$ and encoded by $C$ afterwards.
\end{itemize}
\prettyref{fig:turboSchematic} shows a circuit-type visualization of this definition.
\subsection{Trellis Graphs of Convolutional Codes}
A convolutional code with a specific length is represented naturally by its
trellis graph, which is obtained by unfolding the code-defining finite state machine in the time domain:
Each vertex of the trellis represents the state at a specific point in time, while edges correspond to valid transitions between two subsequent states and exhibit labels with the corresponding input and output bit, respectively. The following description of convolutional codes loosely follows \cite[Section V.C]{Helmling+11MathProgDecoding}, albeit the notation slightly differs.

\begin{figure}\centering

\begin{tikzpicture}[sloped,zero/.style={dashed,->},
                    one/.style={->},
                    vertex/.style={draw,circle,inner sep=.6mm},
                    scale=.8]
\node[vertex] (a0) at (0,0) {$0$};
\node[vertex] (b0) at (1.5,0) {$0$};
\node[vertex] (b2) at (1.5,-2) {$2$};
\draw[zero] (a0) -- node[above] {$0$} (b0);
\draw[one] (a0) -- node[below] {$1$} (b2);
\foreach \pos in {0,...,3} {
  \foreach \state in {0,...,3} {
    \node[vertex] (v\pos\state) at (3 + 1.5*\pos, -\state) {$\state$};
  }
}
\draw[decorate,decoration={brace,mirror}] (-.2,-3.5)-- node[below] {$S_1$} (1.,-3.5);
\draw[decorate,decoration={brace,mirror}] (1.3,-3.5)-- node[below] {$S_2$} (2.5,-3.5);
\foreach \state in {0,...,3} {
  \coordinate (v4\state) at (3 + 6, -\state);
}
\foreach \pos/\next in {0/1,1/2,2/3} {
  \begin{scope}[zero]
    \draw (v\pos0) -- (v\next0);
    \draw (v\pos1) -- (v\next3);
    \draw (v\pos2) -- (v\next1);
    \draw (v\pos3) -- (v\next2);
  \end{scope}
  \begin{scope}[one]
    \draw (v\pos0) -- (v\next2);
    \draw (v\pos1) -- (v\next0);
    \draw (v\pos2) -- (v\next3);
    \draw (v\pos3) -- (v\next1);
  \end{scope}
}
\begin{scope}[gray]
  \clip (6,0) rectangle (8.3,-3);
  \foreach \pos/\next in {3/4} {
    \begin{scope}[zero]
      \draw (v\pos0) -- (v\next0);
      \draw (v\pos1) -- (v\next3);
      \draw (v\pos2) -- (v\next1);
      \draw (v\pos3) -- (v\next2);
    \end{scope}
    \begin{scope}[one]
      \draw (v\pos0) -- (v\next2);
      \draw (v\pos1) -- (v\next0);
      \draw (v\pos2) -- (v\next3);
      \draw (v\pos3) -- (v\next1);
    \end{scope}
  }
\end{scope}
\draw[zero] (b0) -- node[above] {$0$}(v00);
\draw[one] (b0) -- node[above,near start] {$1$}(v02);
\draw[zero] (b2) -- node[auto] {$0$} (v01);
\draw[one] (b2) -- node[below] {$1$} (v03);
\foreach \pos in {1,2,3} {
  \node[vertex,opacity=0.4,dashed] at (0,-\pos) {$\pos$};
}
\foreach \pos in {1,3} {
  \node[vertex,opacity=0.4,dashed] at (1.5,-\pos) {$\pos$};
}
\end{tikzpicture}
\caption{Excerpt from a trellis graph with four states and initial state
$0$. The style of an edge indicates the according information bit, while the labels refer
to the single parity bit.}\label{fig:trellis}
\end{figure}
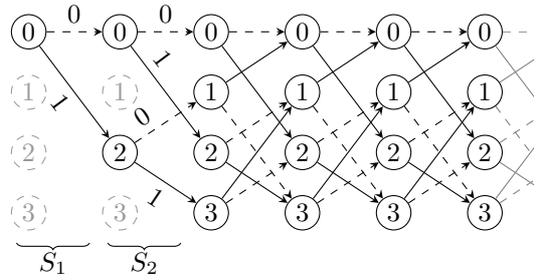

We denote a trellis by $T=(V,E)$ with vertex set $V$ and edge set $E$.
Vertices are indexed by time step and state; \ie, $v_{i,s}$ denotes the vertex corresponding to state $s \in \{0,\dotsc,2^d-1\}$ at time $i \in \{1,\dotsc,k+d+1\}$.
An edge in turn is identified by the time and state of its tail vertex plus its input label, so $e_{i,s,b}$ denotes the edge outgoing from $v_{i,s}$ with input bit $b \in \{0,1\}$.
We call vertical “slices”, \ie, the subgraphs induced by the edges of a single time step, \defn{segments} of the trellis. Formally, the segment at time $i$ is
\begin{align*}
  S_i &= (V_i,E_i) \\
  \text{where}\quad V_i &= \left\{v_{j,s} \in V: j \in \{i,i+1\} \right\}\\
  \text{and}\quad E_i &= \left\{e_{j,s,b}  \in E: j=i\right\}\tp
\end{align*}
Because the initial and final state of the convolutional encoder are fixed, the leading as well as the trailing $d$ segments contain less than $2^d$ vertices.
\prettyref{fig:trellis} shows the first few segments of a trellis with $d=2$.

By construction, the paths from the starting node to the end node in a trellis of a convolutional code $C$ are in one-to-one correspondence with the codewords of $C$: Let $I_j \subset E_j$ and $O_j \subset E_j$ be those edges of $S_j$ whose input label and output label, respectively, is a $1$. The correspondence between a codeword $y \in \F_2^{k+2\cdot d}$ and the according path $P=(e_1,\dotsc,e_{k+d})$ in $T$ is given by
\begin{equation}\begin{aligned}
y_i = 1 &\Leftrightarrow
	\begin{cases}
		e_{k+i} \in I_{k+i}&\text{for }1 \leq i \leq d\\
		e_{i-d} \in O_{i-d}&\text{for }d < i \leq k + 2\cdot d\tk
	\end{cases}
\end{aligned}\label{eq:pathCodewordRelation}\end{equation}
where the first part accounts for the $d$ “input” tail bits that are prepended by convention. From \eqref{eq:pathCodewordRelation}, for each $e \in E$ an index set  $J_C(e)$ can be computed with the property that $e \in P\Rightarrow y_j=1$ for all $j \in J_C(e)$. In our case, $\abs{J_C(e)}$ varies from $0$ (for edges in $S_i$, $i \leq k$, with output label $0$) to $2$ (for edges in $S_i$, $k+1 \leq i \leq k+d$, with both input and output label $1$).

The path-codeword relation can be exploited for maximum likelihood (ML) decoding, if the codewords are transmitted through a memoryless binary-input output-symmetric (MBIOS) channel: Let $\lambda \in \R^{k+2\cdot d}$ be the vector of LLR values of the received signal. If we assign to each edge $e \in E$ the cost
\[c(e) = \sum_{j \in J_C(e)} \lambda_j\tk\]
it can be shown \cite{Feldman03PhD} that the shortest path in $T$ corresponds to the ML codeword.
\subsection{Trellis Representation of Turbo Codes}
For turbo codes, we have two isomorphic trellis graphs, $T^1$ and $T^2$, according to the two component convolutional encoders. Let formally $T = (G^1 \cup G^2, E^1 \cup E^2)$, and by $P=P^1 \circ P^2$ denote the path that consists of $P^1$ in $T^1$ and $P^2$ in $T^2$. Only certain paths, called \defn{agreeable}, actually correspond to codewords; namely, an agreeable path $P_1 \circ P_2=(e^1_1,\dotsc,e^1_{k+d}, e^2_1,\dotsc,e^2_{k+d})$ must obey the $k$ \defn{consistency constraints}
\begin{equation}
	e^1_i \in I^1_i\; \Leftrightarrow \;e^2_{\pi(i)} \in I^2_{\pi(i)}\quad\text{for }i=1,\dotsc,k\label{eq:consistency}
\end{equation}
because both encoders operate on the same information word, only that it is permuted for the second encoder.
Consequently, ML decoding for turbo codes can be formulated as finding the shortest agreeable path in $T$. If an agreeable path contains $e^1_i \in I^1_i$, it must also contain $e^2_{\pi(i)} \in I^2_{\pi(i)}$, and thus $i \in J_C(e)$ for both $e^1_i$ and $e^2_{\pi(i)}$.
To avoid counting the LLR value $\lambda_i$ twice in the objective function, we use the modified cost
\begin{equation}
	c(e) = \sum_{j \in J_C(e)} \hat\lambda_j\text{ with }\hat\lambda_j=
	\begin{cases}
		\frac{\lambda_j}2&\text{if }1\leq j \leq k\tk\\
		\lambda_j&\text{otherwise.}
	\end{cases}\label{eq:edgeCost}
\end{equation}
Then, the ML decoding problem for turbo codes can be stated as the combinatorial optimization problem
\begin{align}
	\text{(TC-ML)}\qquad \min&\sum_{e \in P=P^1\cup P^2}c(e)\label{eq:tcmlObj}\\
	\st\quad&P^1\text{ is a path in }T^1\label{eq:P1path}\\
	&P^2\text{ is a path in }T^2\label{eq:P2path}\\
	&P\text{ is agreeable}\label{eq:Pagree}
\end{align}
The codeword variables $y_i$ can be included into (TC-ML) by the constraints
\begin{equation}
    y_i =
    \begin{cases}
        \sum\limits_{J_C(e) \ni i} \frac{f_e}{2}&\text{for }1 \leq i \leq k\\
        \sum\limits_{J_C(e) \ni i} f_e&\text{for }i > k
    \end{cases}\label{eq:y_in_TCML}
\end{equation}
where the factor $\frac 1 2$ is analogical to \eqref{eq:edgeCost}. However, these variables are purely auxiliary in the LP and thus not needed.

It is straightforward to formulate TC-ML as an integer linear program by introducing a binary flow variable $f_e \in \{0,1\}$ for each $e \in E^1\cup E^2$.
The constraints \eqref{eq:P1path} and \eqref{eq:P2path} can be restated in terms of flow conservation and capacity constraints \cite{Ahuja+93NetworkFlows} which define the path polytopes $\mathcal P^1_{\text{path}}$ and $\mathcal P^2_{\text{path}}$, respectively.
By also transforming \eqref{eq:consistency} and \eqref{eq:tcmlObj}, we obtain
\begin{align}
	\text{(TC-IP)}\qquad\min&\sum_{e \in E^1\cup E^2}c(e)\cdot f_e\\
	\st\quad&f^1\in \mathcal P^1_{\text{path}}\label{eq:P1path-poly}\\
	&f^2\in \mathcal P^2_{\text{path}}\label{eq:P2path-poly}\\
	&\sum_{e \in I^1_i} f_e = \sum_{e \in I^2_{\pi(i)}}f_e&i=1,\dotsc,k\label{eq:Pagree-poly}\\
	&f_e \in \{0,1\}\text{, }e \in E\tp\label{eq:tcip-integrality}
\end{align}

\subsection{Polyhedral Theory Background}
Besides coding theory, this paper requires some bits of polyhedral theory. A \defn{polytope} is the convex hull of a finite number of points: $\mathcal P = \conv{(v_1,\dotsc,v_n)}$. It can be described either by its \defn{vertices} (or \defn{extreme points}), \ie, the unique minimal set fulfilling this defining property, or as the intersection of a finite number of halfspaces: $\mathcal P = \bigcap_{i=1}^m \{x:\,a_i^Tx \leq b_i\}$. An inequality $a^Tx \leq b$ is called \defn{valid} for $\mathcal P$ if it is true for all $x \in \mathcal P$. In that case, the set $F_{a,b} = \{x \in \mathcal P: a^Tx=b\}$ is called the \defn{face induced by} the inequality. For any $r$ satisfying $a^Tr \geq b$ ($a^Tr > b$) we say that the inequality \emph{separates} (\emph{strongly separates}) $r$ from $\calP$.

\section{The LP Relaxation and Conventional Solution Methods}
ML decoding of general linear block codes is known to be $\mathbf{NP}$-hard \cite{Berlekamp+78IntractabilityCoding}.
While the computational complexity of TC-IP is still open, it is widely believed that this problem is $\mathbf{NP}$-hard as well, which would imply that no polynomial-time algorithm can solve TC-IP unless $\mathbf{P}=\mathbf{NP}$\footnote{Note that with state-of-the-art software and prohibitive computational effort, ML turbo decoding \emph{can} be simulated off-line on desktop computers; see \cite{Tanatmis+10NumericalComparison}}. By relaxing \eqref{eq:tcip-integrality} to $f_e \in [0,1]$, we get the LP relaxation (referred to as TC-LP) of the integer program TC-IP, which in contrast can be solved efficiently by the simplex method or interior point approaches \cite{Schrijver86LinearIntegerProg}. Feldman \etal \cite{Feldman+02LPTurboDecoding} were the first to analyze this relaxation and attested it reasonable decoding performance.

A general purpose LP solver, however, does not make use of the combinatorial substructure contained in TC-IP via \eqref{eq:P1path-poly} and \eqref{eq:P2path-poly} and thus wastes some potential of solving the problem more efficiently---while LPs are solvable in polynomial time, they do not scale too well, and the number of variables (about $2\cdot\abs{V}=(k+d)\cdot 2^{d+2}$) and constraints ($\abs{V}+k$) in TC-LP is very large (practical values of $d$ range roughly from $3$ to $8$).

Note that without the consistency constraints \eqref{eq:Pagree-poly}, we could solve TC-LP by simply computing shortest paths in both trellis graphs, which is possible in time $\mathcal O(k+d)$, even in the presence of negative weights, because the graphs are acyclic \cite{Cormen+01Algorithms}. A popular approach for solving  optimization problems that comprise “easy” subproblem plus some “complicating” additional constraints is to solve the Lagrangian dual \cite{NemhauserWolsey88} by subgradient optimization. If we define $g_i(f) = \sum_{e\in I_i^1}f_e-\sum_{e\in I_{\pi(i)}^2}f_e$, the constraints \eqref{eq:Pagree-poly} can be compactly rewritten as
\begin{equation}g_i(f)=0\quad\text{for }i=1,\dotsc,k\tp\label{eq:defg_i}\end{equation}
The \defn{Lagrangian relaxation with multiplier $\mu \in \R^k$} is defined as
\begin{align}
\text{(TC-LR)}\quad z(\mu)=\min&\sum_{e \in E^1\cup E^2}c(e)\cdot f_e + \sum_{i=1}^k\mu_k\cdot g_i(f)\\
	\st\quad&f^1\in \mathcal P^1_{\text{path}}\\
	&f^2\in \mathcal P^2_{\text{path}}\\
	&f_e \in \{0,1\}\text{, }e \in E
\end{align}
For all $\mu \in \R^k$, the objective value of TC-LR is smaller or equal to that of TC-LP. The \defn{Lagrangian dual problem} is to find multipliers $\mu$ that maximize this objective, thus minimizing the gap to the LP solution.
It can be shown that in the optimal case both values coincide.
Note that the feasible region of TC-LR is the combined path polytope of both $T_1$ and $T_2$, so it can be solved by a shortest path routine in both trellises with modified costs, and the integrality condition on $f$ is fulfilled automatically.
Applying Lagrangian relaxation to turbo decoding was already proposed by Feldman \etal~\cite{Feldman+02LPTurboDecoding} and further elaborated by Tanatmis \etal~\cite{Tanatmis+10Lagrangian}; the latter reference combines the approach with a heuristic to tighten the integrality gap between TC-LP and TC-IP.

The Lagrangian dual is typically solved by a subgradient algorithm that iteratively adjusts the multipliers $\mu$, converging (under some mild conditions) to the optimal value \cite{NemhauserWolsey88}. However, the convergence is often slow in practice and the limit is not guaranteed to be ever reached exactly. Additionally, the dual only informs us about the objective value; recovering the actual solution of the problem requires additional work. In summary, subgradient algorithms suffer from three major flaws. The main result of this paper is an alternative algorithm which exhibits none of these.

\section{An Equivalent Problem in Constraints Space}\label{sec:theory}
Like Lagrangian dualization, our algorithm also uses a relaxed formulation of TC-IP with modified objective function that resembles TC-LR. However, via geometric interpretation of the image of the path polytope in the “constraints space”, as defined below, the exact LP solution is found in finitely many steps.

\subsection{The Image Polytope \texorpdfstring{$\calQ$}{Q}}
Let $\Ppath = \Ppath^1 \times \Ppath^2$ be the feasible region of TC-LR. We define the map
\begin{align}
	\mathfrak D: \Ppath &\rightarrow \R^{k+1}\\
	f & \mapsto (g_1(f),\dotsc,g_k(f),c(f))^T
\end{align}
where $c(f) = \sum_{e \in E^1\cup E^2}c(e)\cdot f_e$ is a short hand for the objective function value of TC-LP. For a path $f$, the first $k$ coordinates $v_i, i=1,\dotsc,k$, of $v = \mathfrak D(f)$ tell if and how the condition $g_i(f)=0$ is violated, while the last coordinate $v_{k+1}$ equals the cost of $f$. Let $\calQ = \mathfrak D(\Ppath)$ be the image of the path polytope under $\mathfrak D$. The following results are immediate:
\begin{lemma}\label{lemma:PQcorrespondence}~\hfill
	\begin{enumerate}
		\item $\calQ$ is a polytope.
		\item If $f$ represents an agreeable path in $T$, then $\mathfrak D(f)$ is located on the $(k+1)$st axis (henceforth called $c$-axis or $A_c$).
		\item If $v$ is a vertex of $\calQ$ and $v = \mathfrak D(f)$ for some $f \in \Ppath$, then $f$ is also a vertex of $\Ppath$.
	\end{enumerate}
\end{lemma}
In the situation that $v = \mathfrak D(f)$ we will also write $f = \mathfrak D^{-1}(v)$ with the meaning that $f$ is \emph{any} preimage of $v$, which need not be unique.

We consider the auxiliary problem
\begin{align}
	\text{(TC-LP$_\calQ$)}\quad z^\calQ_\text{LP}=&\min\,v_{k+1}\\
	\st\quad&v \in \calQ\\
	&v \in A_c\label{eq:TCLPQonAxis}
\end{align}
the solution of which is the lower “piercing point” of the axis $A_c$ through $\calQ$. Note that due to \eqref{eq:TCLPQonAxis}, $k$ of the $k+1$ variables in TC-LP($\calQ$) are fixed to zero, thus the problem is in a sense one-dimensional, the feasible region being the (one-dimensional) projection of $\calQ$ onto $A_c$. Nevertheless, the following theroem shows that TC-LP$_\calQ$ and TC-LP are essentially equivalent.
\begin{theorem}\label{thm:solutionLink}
	Let $v_{\text{LP}}$ be an optimal solution of TC-LP$_\calQ$ with objective value $z^\calQ_{\text{LP}}$ and $f_{\text{LP}}=\mathfrak D^{-1}(v_{\text{LP}}) \in \Ppath$ the corresponding flow. Then $z^\calQ_{\text{LP}} = z_{\text{LP}}$, the optimal objective value of TC-LP, and $f_{\text{LP}}$ is an optimal solution of TC-LP.
\end{theorem}
\begin{IEEEproof}
First we show $z^\calQ_{\text{LP}} \leq z_{\text{LP}}$. Let $f_{\text{LP}}$ be an optimal solution of TC-LP with cost $c(f_{\text{LP}})=z_{\text{LP}}$. Then $\mathfrak D(f_{\text{LP}}) = (0,\dotsc,0,z_{\text{LP}})$ by definition of $\mathfrak D$, since $f_{\text{LP}}$ is feasible and thus $g_1(f_{\text{LP}}) = \dotsm = g_k(f_{\text{LP}}) = 0$. Hence $\mathfrak D(f_{\text{LP}}) \in A_c \cap \calQ$ with $\mathfrak D(f_{\text{LP}})_{k+1} = z_{\text{LP}}$, from which it follows that $z^\calQ_{\text{LP}} \leq z_{\text{LP}}$.

If we assume on the other hand that $z^\calQ_{\text{LP}} < z_{\text{LP}}$, there must be a $v \in A_c \cap \calQ$ such that $v_{k+1} < z_{\text{LP}}$. By definition of $\mathfrak D$ this implies the existence of a flow $f=\mathfrak D^{-1}(v)$ with $g_1(f) = \dotsm = g_k(f) = 0$, hence a feasible one, and $c(f) = v_{k+1} < z_{\text{LP}}$, contradicting optimality of $z_{\text{LP}}$.
\end{IEEEproof}
While we do not have an explicit representation of $\calQ$---by means of either vertices or inequalities---at hand, we can easily minimize linear functionals over $\calQ$:
\begin{observation}\label{obs:LPQisTCWS}
The problem
\begin{align}
	\text{(LP$_\calQ$)}\qquad\min\,& \gamma^T v\\
	\st\quad&v \in \calQ
\end{align}
can be solved by first computing an optimal solution $f^\ast$ of the weighted sum problem
\begin{align}
	\text{(TC-WS)}\qquad\min\,&\sum_{i=1}^k \gamma_i\cdot g_i(f) + \gamma_{k+1}\cdot c(f)\\
	\st\quad&f \in \Ppath
\end{align}
and then taking the image of $f^\ast$ under $\mathfrak D$. As noted before, this can be achieved within running time $\O(n)$.
\end{observation}
Note that TC-WS is closely related to TC-LR: as long as $\gamma_{k+1} \neq 0$, we get the same problem by setting $\mu_i = \frac{\gamma_i}{\gamma_{k+1}}$ in TC-LR.
\subsection{Solving TC-\texorpdfstring{LP$_\calQ$}{LP-Q} with Nearest Point Calculations}
Our algorithm solves TC-LP$_\calQ$ by a series of nearest point computations between $\calQ$ and reference points $r^i$ on $A_c$, the last of which gives a face of $\calQ$ containing the optimal solution $v_{\text{LP}}$.

For each $r \in \R^{k+1}$, we denote by 
\begin{equation}\op{NP}(r)=\argmin_{v \in \calQ} \norm{v - r}_2\end{equation}
the nearest point to $r$ in $\calQ$ with respect to Euclidean norm and define
\begin{align}
	&a(r) = r-\op{NP}(r)\\
	&b(r)=a(r)^T\op{NP}(r)\tp
\end{align}
The following well-known result will be used frequently below.
\begin{lemma}\label{lemma:mzfpSeparation}
The inequality
\begin{equation}
a(r)^T v \leq b(r)\label{eq:NPseparating}
\end{equation}
is valid for $\calQ$ and induces a face containing $\op{NP}(r)$, which we call $\op{NF}(r)$. If $r \notin \calQ$, \eqref{eq:NPseparating} strongly separates $r$ from $\calQ$.
\end{lemma}
The following theorem is the foundation of our algorithm.
\begin{theorem}\label{thm:mzfpMainTheorem}
	There exists an $\varepsilon>0$ such that for all $r$ inside the open line segment $\left(v_{\text{LP}}, v_{\text{LP}}- (0,\dotsc,0,\varepsilon)^T\right)$ the condition
	\begin{equation}v_{\text{LP}} \in \op{NF}(r)\end{equation}
	holds.
\end{theorem}
Our constructive proof of \prettyref{thm:mzfpMainTheorem} shows how find a point inside the interval mentioned in the theorem. The outline is as follows: At first, start with a reference point $r \in A_c$ that is guaranteed to be located below $v_{\text{LP}}$. Then, we iteratively compute $\op{NF}(r)$ and update $r$ to be the intersection of $A_c$ with the hyperplane defining $\op{NF}(r)$. The following lemmas show that this procedure is valid and finite.

The first result is that the hyperplane defining $\op{NF}(r)$ is always oriented “downwards”.
\begin{lemma}\label{lemma:mzfpIneqOrientation}
Let $r=(0,\dotsc,0,\rho)^T$ with $\rho < z^\calQ_{\text{LP}}$ and let $a(r)^T v \leq b(r)$ be the inequality defined in \eqref{eq:NPseparating}. Then, \[a(r)_{k+1} < 0\tp\]
\end{lemma}
\begin{IEEEproof}
Assuming $a(r)_{k+1} \geq 0$, we obtain
$a(r)^T v_{\text{LP}} = a(r)_{k+1} z^\calQ_{\text{LP}} \geq a(r)_{k+1} \rho = a(r)^T r > b(r)$, which contradicts $v_{\text{LP}} \in \calQ$ by \prettyref{lemma:mzfpSeparation}.
Note that the equalities hold because both $v_{\text{LP}}$ and $r$
are elements of $A_c$, the first inequality stems from the assumptions on $a(r)_{k+1}$ and $\rho$, and the second follows from \prettyref{lemma:mzfpSeparation}.
\end{IEEEproof}
Next we show that updating $r$ leads to a different nearest face, unless we have arrived at the optimal solution.
\begin{lemma}\label{lemma:mzfpNotSameFace}
Under the same assumptions as in \prettyref{lemma:mzfpIneqOrientation}, let $s=(0,\dotsc,0,s_{k+1})$ with $s_{k+1} = \frac{b(r)}{a(r)_{k+1}}$ be the point where the separating
hyperplane and $A_c$ intersect. If $\op{NF}(r) = \op{NF}(s)$, then $s = v_{\text{LP}}$.
\end{lemma}
\begin{IEEEproof}
We use contraposition to show that $s \neq v_{\text{LP}}$ implies
$\op{NF}(r) \neq \op{NF}(s)$, so assume
$s \neq v_{\text{LP}}$. We know that $a(r)^T v \leq b(r)$ is valid
for $\calQ$ and $a(r)^T s = a(r)_{k+1} s_{k+1} = b(r)$ by construction. This implies that $s \notin \calQ$; otherwise we would have $s = v_{\text{LP}}$
because for all $\zeta < s_{k+1}$, $a(r)^T (\zeta e_{k+1}) > b(r)$, so $s$ would
really be the lowest point on $A_c$ that is also in $\calQ$ and thus
optimal.

It follows that $y_N = \op{NP}(s) \neq s$. Since
$y_N \in \calQ$ and $a(r)^T v \leq b(r)$ is valid for $\calQ$,
we have $a(r)^T y_N \leq b(r)$.

\textbf{Case 1:} $a(r)^T y_N < b(r)$. Then $y_N \notin \op{NF}(r)$,
but $y_N \in \op{NF}(s)$ by definition, which proves the claim for this case.

\textbf{Case 2:} $a(r)^T y_N = b(r)$.
From $a(r)^T r > b^r$ and $a(r)^T s = b(r)$ we obtain
\begin{align}
& a(r)^T r > a(r)^T s \\
\Rightarrow\, &a(r)^T (r-s) > 0\\
\Rightarrow\, &a(r)_{k+1} (r_{k+1} - z_{k+1}) > 0\\
\Rightarrow \,&r_{k+1} < s_{k+1}\tk\label{eq:zdgreaterrd}
\end{align}
where we have used again $a(r)_{k+1}<0$ and the fact that $r, s \in A_c$.

Applying \prettyref{lemma:mzfpIneqOrientation}
to $s$ as reference point we obtain $a(s)_{k+1} = (s-y_N)_{k+1} < 0$, hence
\begin{align}&(y_N)_{k+1} > s_{k+1}\\
\Rightarrow\,&(y_N)_{k+1}(s_{k+1}-r_{k+1}) > s_{k+1}(s_{k+1}-r_{k+1})&\text{by \eqref{eq:zdgreaterrd}}\\
\Rightarrow\,&y_N^T(s-r) > s^T(s-r)\\
\Rightarrow\,&y_N^Ts - y_N^Tr + s^Tr - s^Ts > 0\label{eq:mzfpProofBreak}
\intertext{
Plugging the definitions into $a(r)^T y_N = b(r) = a(r)^Ts$ yields
$(r-x_N)^Ty_N = (r-\op{NP}(r))^T s$ or $r^T s - r^Ty_N = \op{NP}(r)^Ts - \op{NP}(r)^Ty_N$. Using this we continue from \eqref{eq:mzfpProofBreak} with
}
\Rightarrow\,&y_N^Ts + \op{NP}(r)^Ts - \op{NP}(r)^Ty_N - s^Ts > 0\\
\Rightarrow\,&\op{NP}(r)^T(s-y_N) > s^T(s-y_N)\\
\Rightarrow\,&a(s)^T \op{NP}(r) > a(s)^T s > b(s)
\end{align}
Thus, $\op{NP}(r) \notin \op{NF}(s) = \{ v \in \calQ:\, a(s)^Tv \leq b(s)\}$, but $\op{NP}(r) \in \op{NF}(r)$ by definition, so those faces must differ.
\end{IEEEproof}
Now we show the auxiliary result that if two inequalities induce the same face, then also every convex combination of them does.
\begin{lemma}
\label{lemma:faceDefiningInequalities}
Let $\mathcal P$ be a polytope, $x^1, x^2 \in \mathcal P$, and
$r^1, r^2 \notin \mathcal P$. If the inequalities
\[H^1: (r^1-x^1)^Tx \leq (r^1-x^1)^T x^1\]
and
\[H^2: (r^2-x^2)^Tx \leq (r^2-x^2)^T x^2\]
both induce the same face $F$ of $\mathcal P$, then also
\[\bar H: (\bar r - \bar x)^T x \leq (\bar r - \bar x)^T \bar x\]
with $\bar r = \lambda r^1 + (1-\lambda)r^2$, $\bar x = \lambda x^2 + (1-\lambda) x^2$, $0 \leq \lambda \leq 1$,
is valid and induces $F$.
\end{lemma}
\begin{IEEEproof}
We first show that $\bar H$ is valid. For $x \in \mathcal P$
\begin{align*}
(\bar r - \bar x)^Tx &= \lambda(r^1-x^1)^Tx + (1-\lambda)(r^2-x^2)^Tx\\
&\leq \lambda(r^1-x^1)^Tx^1 + (1-\lambda)(r^2-x^2)^Tx^2\\
&=\left(\lambda(r^1-x^1)+(1-\lambda)(r^2-x^2)\right)^T\\
&\quad \left(\lambda x^1+(1-\lambda)x^2\right)\\
&=(\bar r - \bar x)^T \bar x
\tk
\end{align*}
where we have used the fact that $H^1$ is satisfied with equality for $x=x^2$ and vice versa because of the assumptions. Since we have shown that $\bar H$
is valid, it must induce a face $\bar F$. It remains to show that $F=\bar F$.

``$F \subseteq \bar F$'': $x \in F$ fulfills both $H^1$ and $H^2$ with equality, so we can carry out the above calculation with a ``$=$'' in the second line to conclude $x \in \bar F$.

``$\bar F \subseteq F$'': Let $x \in \bar F$ and assume $x \notin F$, which
implies $(r^i-x^i)^Tx < (r^i-x^i)^Tx^i$ for $i \in \{1,2\}$. Then
$\lambda(r^1-x^1)^Tx^1+(1-\lambda)(r^2-x^2)^Tx^2
> \lambda(r^1-x^1)^Tx + (1-\lambda)(r^2-x^2)^Tx
= (\bar r - \bar x)^T x
= (\bar r - \bar x)^T\bar x
= \lambda(r^1-x^1)^Tx^1+(1-\lambda)(r^2-x^2)^Tx^2$, which is a contradiction.
\end{IEEEproof}
The above lemma is used to show that the part of $A_c$ that lies below $v_{\text{LP}}$ dissects into intervals such that reference points within one interval yield the same face of $\calQ$.
\begin{lemma}\label{lemma:mzfpFaceIntervals}
If $r^1 < r^2 < x^\ast_d$ and $\op{NF}(r^1) = \op{NF}(r^2)$, then $\op{NF}(r) = \op{NF}(r^1)$ for all $r \in [r^1,r^2]$.
\end{lemma}
\begin{IEEEproof}
Let $v^i = \op{NP}(r^i)$ for $i \in \{1,2\}$. By \prettyref{lemma:faceDefiningInequalities}, for each $\lambda \in (0,1)$ 
and $\bar r = \lambda r^1 + (1-\lambda)r^2$, $\bar v = \lambda v^1 + (1-\lambda)v^2$, it holds
\[\{v \in \calQ: (\bar r - \bar v)^T v = (\bar r - \bar v)^T\bar v\} = \op{NF}(r^1)\tk\]
and applying the converse statement from \prettyref{lemma:mzfpSeparation}
follows $\bar v = \op{NP}(\bar r)$, so $\op{NF}(\bar r) = \op{NF}(r^1)$ as claimed.
\end{IEEEproof}
Now we have alle the ingredients at hand to prove our theorem.
\begin{IEEEproof}[Proof of \prettyref{thm:mzfpMainTheorem}]
First we show that there exists at least one $r$ with the
desired properties.

Choose some arbitrary $r^0 \in A_c$ with
$r^0_{k+1} < z^\calQ_{\text{LP}}$ (thus $r^0 \notin \calQ$). If $v_{\text{LP}} \in \op{NF}(r^0)$, we are done. Otherwise,
\prettyref{lemma:mzfpNotSameFace} tells us how to find an
$r^1$ with $r^1_{k+1} > r^0_{k+1}$ such that $\op{NF}(r^1) \neq \op{NF}(r^0)$. Iterating this argument and assuming that $v_{\text{LP}}$ is never contained in the induced face results in a sequence $(r^i)_i$ with $r^{i+1}_{k+1} > r^i_{k+1}$ for all $i$. Because of \prettyref{lemma:mzfpFaceIntervals}, 
$\op{NF}(r^{i+1}) \neq \op{NF}(r^i)$ implies
$\op{NF}(r^{i+1}) \neq \op{NF}(r^l)$ for all
$0 \leq l < i+1$, so that all $\op{NF}(r^i)$ are distinct.
But since there are only finitely many faces of $\calQ$, this
can not be true, so eventually there must be an ${i^\ast}$ such that
$v_{\text{LP}} \in \op{NF}(r^{i^\ast})$.

Now let $r^\ast \in A_c$ be any such point whose existence we
have just proven, $v_N = \op{NP}(r^\ast)$ and $\lambda \in (0,1]$. Let $\bar r = \lambda r^\ast + (1-\lambda)v_{\text{LP}}$ and $\bar v = \lambda v_N + (1-\lambda) v_{\text{LP}}$. We use similar arguments as in
the proof of \prettyref{lemma:faceDefiningInequalities}
to show that
\begin{equation}(\bar r - \bar v)^T v \leq (\bar r - \bar v)^T\bar v\label{eq:mzfpFaceStar}\end{equation}
induces $\op{NF}(r^\ast)$.

For $v \in \calQ$,
\begin{align*}
(\bar r - \bar v)^T v & = (\lambda(r^\ast-v_N)+(1-\lambda)(v_{\text{LP}} -v_{\text{LP}}))^Tv\\
& = \lambda(r^\ast-v_N)^Tv\\
& \leq \lambda(r^\ast-v_N)^Tv_N\\
& = \lambda(\lambda(r^\ast-v_N)^Tv_N+(1-\lambda)(r^\ast-v_N)^Tv_{\text{LP}})\\
& = \lambda(r^\ast-v_N)^T(\lambda v_N + (1-\lambda)v_{\text{LP}})\\
& = (\bar r - \bar v)^T \bar v\tp
\end{align*}
So the inequality is valid, and since again for $v \in \op{NF}( r^\ast)$ equality holds in the third line, we know that the face
$\bar F$ induced by \eqref{eq:mzfpFaceStar} contains $\op{NF}( r^\ast)$.

Now let $v \in \bar F$, \ie, $(\bar r - \bar x)^Tv = (\bar r - \bar x)^T\bar x$. From the above equations we conclude $\lambda(r^\ast-v_N)^Tv = \lambda(r^\ast-v_N)^Tv_N$, and because $\lambda > 0$ this implies
$v \in \op{NF}(r^\ast)$.

Because the above holds for any $0 < \lambda \leq 1$, we can choose
$\bar r$ arbitrarily close to $v_{\text{LP}}$ on $A_c$, which completes the proof.
\end{IEEEproof}
An illustration of the process in two dimensions is given in \prettyref{fig:mzfpExample}.
\begin{figure*}
\centering
\subcaptionbox{Step i: $v^i=\mathfrak D(f^i)$ is found as nearest point to some reference point $r^{i-1}$. The intersection of the separating hyperplane with the axis $A_c$, $r^1$, will be the reference point of the next iteration.}[0.45\textwidth]
{
	\begin{tikzpicture}[scale=.6]
	\coordinate (v1) at (-4,-3.2);
	\coordinate (v2) at (-2.5, -3);
	\coordinate (v3) at (-0.5,-1);
	\coordinate (v4) at (1.5,3);
	\coordinate (v5) at (-4,3);
    \draw (v1) -- (v2) -- (v3) -- (v4);
	\begin{scope}[draw=green!70!black,text=green!70!black]
		\node[below] at (v2) {$v^i$};
		\draw[thick] (-4.5,-4) -- (0.5, -1.5);
		\draw[dashed] (-1.5,-5) -- (v2);
		\node[dot,label=right:$r^i$] at (0,-1.75) {};
	\end{scope}
	\fill[fill=red!10] (v1) -- (v2) -- (v3) -- (v4) -- (v5) -- cycle;

	\node at (-1.5,1.5) {$\mathcal Q$};
	\foreach \i in {2,...,3}
	{
	 \node[dot] at (v\i) {};
	}
	\draw[->,thick] (0,-5) -- (0,3.5) node[right] {$A_c$};
	\foreach \y in {-4,...,2}
	{\draw (.1,\y)--(-.1,\y);}
	\end{tikzpicture}
}\quad
\subcaptionbox{Step $i+1$: Note that the induced face of $\calQ$ here is a facet, while it was a $0$-dimensional face in step $i$.\label{fig:mzfpExampleB}}[0.45\textwidth]
{
	\begin{tikzpicture}[scale=.6]
	\coordinate (v1) at (-4,-3.2);
	\coordinate (v2) at (-2.5, -3);
	\coordinate (v3) at (-0.5,-1);
	\coordinate (v4) at (1.5,3);
	\coordinate (v5) at (-4,3);
	\fill[fill=red!10] (v1) -- (v2) -- (v3) -- (v4) -- (v5) -- cycle;
	\draw (v1) -- (v2) -- (v3) -- (v4);
	\node[below,text=green!70!black] at (v2) {$v^i$};
	\begin{scope}[draw=blue,text=blue]
		\node[dot,label=left:$v^{i+1}$] (xn) at (-0.625, -1.125) {};
		\node[dot,label=right:$r^i$] (r) at (0,-1.75) {};
		\node[dot,label=right:$r^{i+1}$] at (0,-.5) {};
		\draw (r) -- (xn);
		\draw[thick] (-3.5,-4) -- (1,.5);
	\end{scope}
	\node at (-1.5,1.5) {$\mathcal Q$};
	\foreach \i in {2,...,3}
	{
	 \node[dot] at (v\i) {};
	}
	\draw[->,thick] (0,-5) -- (0,3.5) node[right] {$A_c$};
	\foreach \y in {-4,...,2}
	{\draw (.1,\y)--(-.1,\y);}
	\end{tikzpicture}
}\\
\subcaptionbox{Step $i+2$ (zoomed in): The facet induced in this step intersects $A_c$ at $v_{\text{LP}}$, but the algorithm can not yet detect this.
\label{fig:mzfpExampleC}}[0.45\textwidth]
{
	\begin{tikzpicture}[scale=1.4]
	\coordinate (v2) at (-1, -1.5);
	\coordinate (v3) at (-0.5,-1);
	\coordinate (v4) at (0.75,1.5);
	\coordinate (v5) at (-1,1.5);
	\fill[fill=red!10] (v2) -- (v3) -- (v4) -- (v5) -- cycle;
	\draw (v2) -- (v3) -- (v4);
	\begin{scope}[draw=red,text=red]
		\node[dot,label=right:$r^{i+1}$] (r) at (0,-.5) {};
		\coordinate (xn) at (-.2,-.4) {};
		\draw (r) -- (xn);		
		\draw[thick] (-0.75,-1.5) -- (.5,1);	
		\node[dot,label=above left:${r^{i+2}=v_{\text{LP}}}$] at (0,0) {};
		\node[dot,label=left:$v^{i+2}$] at (xn) {};
	\end{scope}
	\node[dot] at(-0.625, -1.125) {};
	\node[blue,left] at(-0.625, -1.125) {$v^{i+1}$};
	\node at (-.5,1) {$\calQ$};
	\foreach \i in {3}
	{
	 \node[dot] at (v\i) {};
	}
	\draw[->,thick] (0,-1.5) -- (0,2) node[right] {$A_c$};
	\foreach \y in {-1,...,1}
	{\draw (.1,\y)--(-.1,\y);}
	\end{tikzpicture}
}\quad
\subcaptionbox{Step $i+3$: Optimality is detected by $v^{i+3}=r^{i+2}$. The solution $\mathfrak D^{-1}(v_{\text{LP}})$ is returned.
\label{fig:mzfpExampleD}}[0.45\textwidth]
{
	\begin{tikzpicture}[scale=1.4]
	\coordinate (v2) at (-1, -1.5);
	\coordinate (v3) at (-0.5,-1);
	\coordinate (v4) at (0.75,1.5);
	\coordinate (v5) at (-1,1.5);
	\fill[fill=red!10] (v2) -- (v3) -- (v4) -- (v5) -- cycle;
	\draw (v2) -- (v3) -- (v4);
	\node[dot] at (xn) {};
	\node[left,text=red] at (xn) {$v^{i+2}$};
	\begin{scope}[text=cyan]
		\node[dot,label=above left:${r^{i+2}=v^{i+3}=v_{\text{LP}}}$] at (0,0) {};
	\end{scope}
	\node[dot] at(-0.625, -1.125) {};
	\node[blue,left] at(-0.625, -1.125) {$v^{i+1}$};
	\node at (-.5,1) {$\mathcal Q$};
	\foreach \i in {3}
	{
	 \node[dot] at (v\i) {};
	}
	\draw[->,thick] (0,-1.5) -- (0,2) node[right] {$A_c$};
	\foreach \y in {-1,...,1}
	{\draw (.1,\y)--(-.1,\y);}
	\end{tikzpicture}
}
\caption{Schematic execution of \prettyref{alg:mzfpNP} in image space}
\label{fig:mzfpExample}
\end{figure*}
\subsection{Solving the Nearest Point Problems}\label{sec:npp}
It remains to show how to solve the nearest point problems arising in the discussion above. To that end, we utilize an algorithm by Wolfe \cite{Wolfe76NearestPoint} that finds in a polytope the point with minimum Euclidean norm. Wolfe's algorithm elaborates on a set of vertices of the polytope that are obtained via minimization of linear objective functions. In our situation, this means that LP$_\calQ$ has to be solved repeatedly, which by \prettyref{obs:LPQisTCWS} boils down to the linear-time solvable weighted sum shortest path problem TC-WS. Note that by subtracting $r$ from the results of LP$_\calQ$ and adding $r$ to the final result, the algorithm can be used to calculate the minimum distance between $\calQ$ and $r$ also in the case $r \neq 0$.

The algorithm in \cite{Wolfe76NearestPoint} maintains in each iteration a subset $P$ of the vertex set $V(\calQ)$ and a point $x$ such that $x = \op{NP}(\aff(P))$ lies in the relative interior of $\conv(P)$, where $\aff(P)$ is the affine hull of $P$. Such a set is called a \defn{corral}, and we denote the nearest point in $\aff(P)$ by $v^{\aff}_P$.

Initially $P=\{v_0\}$ for an arbitrary vertex $v_0$ and $x=v_0$. Note that then $v^{\aff}_P=v_0$ and $P$ is indeed a corral. Then the following is executed iteratively (we explain afterwards how the computations are actually carried out):
\begin{enumerate}
  \item\label{item:npa_step1} Solve $p = \argmin_{v \in \calQ}(x^Tv)$.
  \item If $p=0$ ($0$ is optimal) or $x^T p = x^T x$ ($x$ is optimal), stop. Otherwise, set $P := P \cup \{p\}$ and compute $y = v^{\aff}_P$.
  \item\label{item:npa_minorEntry} If $y$ is in the relative interior of $\conv(P)$, $P$ is a corral. Set $x := y$ and continue at \ref{item:npa_step1}).
  \item Determine $z \in \conv(P) \cap \conv\{x, y\}$ with minimum distance to $y$; $z$ will be a boundary point of $\conv(P)$.
  \item Remove from $P$ some point that is not on the smallest face of $\conv(P)$ containing $z$, set $x := z$, and continue at \ref{item:npa_minorEntry}).
\end{enumerate}
The algorithm will eventually find a corral $P$ such that the nearest point of $\calQ$ equals $v^{\aff}_P$.

The computations in each step are performed as follows:
\begin{enumerate}
  \item This matches the solution of TC-WS.
  \item If we interchangeably use the symbol $P$ for both the set of points and the matrix that contains the elements of $P$ as columns, every $v \in \aff(P)$ can be characterized by some $\lambda \in \R^{\abs{P}}$ such that $v = P\lambda$ and $e^T\lambda=1$. Thus, the subproblem of determining $v^{\aff}_P$ can be written as
\begin{align}
    \min\;&\norm{P\lambda}_2^2 = \lambda^TP^TP\lambda\\
    \text{\st}\quad&e^T \lambda = 1\tp
\end{align}
It can be shown \cite{Wolfe76NearestPoint} that this is equivalent to solving the system of linear equations
\begin{equation}\begin{aligned}
    (e e^T + P^TP)\mu = e\\
    \lambda = \frac 1 {\norm{\mu}_1} \mu
\end{aligned}\label{eq:npa_linearEq}
\end{equation}
As an efficient method to solve \eqref{eq:npa_linearEq}, Wolfe suggests to maintain an upper triangular matrix $R$ such that $R^T R = ee^T + P^TP$. Then the solution $\mu$ can be found by first solving $R^T \bar \mu = e$ for $\bar \mu$ and then $R\mu = \bar \mu$ for $\mu$; both can be done by a simple backward substitution. When $P$ changes, $R$ can be updated relatively easily without the necessity of a complete recomputation \cite{Wolfe76NearestPoint}.
  \item $y$ is in the relative interior of $\conv(P)$ if and only if all coefficients $\lambda_i$ in the convex representation of $y$ satisfy $\lambda_i > 0$.
  \item By construction $x \in \conv(P)$. Let $x=\sum_{v \in P} \lambda_v v$ and $y=\sum_{v \in P}\mu_v v$, where $\sum_{v \in P} \lambda_v = \sum_{v \in P} \mu_v = 1$, but $\mu_p \leq 0$ for at least one $p$. The goal can then be restated as finding the minimal $\theta \in [0,1]$ such that $z_\theta = \theta x + (1-\theta)y \in \conv(P)$. Substituting the above expressions yields
  \[z_\theta = \sum_{v \in P} \left( \theta \lambda_v + (1-\theta) \mu_v \right)v\tk\]
  and the condition is that all coefficients are nonnegative. Thus, for all $v$ with $\mu_v \leq 0$,
  \[ \theta \geq \frac{\mu_p}{\mu_p-v_p} \]
  must hold. In summary, $\theta$ can be computed by the rule
  \[ \theta = \min\left\{1, \max\left\{\frac{\mu_p}{\mu_p-v_p}:\, \mu_p < 0\right\} \right\}\tp\]
  \item A point not contained in the smallest face of $\conv(P)$ containing $z$ is not needed for the convex description of $z=\sum_{v \in P} \lambda_v v$; thus it can be identified by $\lambda_v = 0$.
\end{enumerate}

\subsection{Recovering the Optimal Flow and Pseudocodeword}
So far we have shown how to compute the optimal \emph{objective value}, but not the LP \emph{solution}, i.\,e.\ the flow $f_\text{LP} \in \Ppath$ and the (pseudo)codeword $y$. The algorithm yields its solution $v_\text{LP}$ by means of a convex combination of extreme points of $\calQ$:
\[v_\text{LP} = \sum_{i=1}^t \lambda_i v_i ,\quad\lambda_i\geq 0,\quad\sum_{i=1}^t \lambda_i=1\tp\]
During its execution the preimage paths $f_i=\mathfrak D^{-1}(v_i)$ can be stored alongside with the $v_i$. Then, the LP-optimal flow $f_\text{LP}$ is obtained by summing up the paths with the same weight coefficients $\lambda$, \ie,
\[f_\text{LP}=\sum_{i=1}^t \lambda_i  \mathfrak D^{-1}(v_i)\tp\]
In order to get the corresponding pseudocodeword, a simple computation based on \eqref{eq:y_in_TCML} suffices.

For most applications, however, the values of $y$ are of interest only in the case that the decoder has found a valid codeword, \ie, $t=1$ in the above sum. In such a case, the most recent solution of (TC-WS) is an agreeable path that immediately gives the codeword. No intermediate paths have to be stored, which can save a substantial amount of space and running time.

\subsection{Efficient Reference Point Updates}
As suggested by the proof of \prettyref{thm:mzfpMainTheorem}, the nearest point algorithm is run iteratively, and between two runs the $k+1$st component of $r$ is increased by means of the rule $r_{k+1} = \frac{b(r)}{a(r)_{k+1}}$. This section describes how some information from the previous iteration can be re-used to provide an efficient warm start for the next nearest point run.

Assume that in iteration $i$ the point $\op{NP}(r^i) = v^{i+1}$ has been found, inducing the face $\op{NF}(r^i)$ defined by $a(r^i)^T v \leq b(r^i)$ of $\calQ$. Recall that NPA internally computes the minimum $l_2$ norm of $\calQ - r^i$. Thus, it outputs $\bar v^{i+1} = v^{i+1} - r^i$ as the convex combination of $t \leq k+1$ points $\bar v_j = v_j-r^i$, all of which are located on the corresponding face $\hat{\op{NF}}(r^i)$ of $\calQ - r^i$:
\[\bar v^{i+1} =  v^{i+1} -r^i = \sum_{j=1}^t \lambda_j (v_j-r^i) = \sum_{j=1}^t \lambda_j \bar v_j\]

In the subsequent nearest point calculation, the norm of $\calQ-r^{i+1}$ is minimized. Obviously $\hat{\op{NF}}(r^i)$ corresponds to a face $\hat{\op{NF}}(r^{i+1})$ of $\calQ-r^{i+1}$, and we can initialize the algorithm with that face by simply adding $r^i-r^{i+1}$ to $\bar v$ and each $\bar v_j$, $j=1,\dotsc,t$, which yields
\[\bar v^{i+1} + r^i-r^{i+1} = v^{i+1}-r^{i+1} = \sum_{j=1}^t \lambda_j (v_j-r^{i+1})\]
and all $v_j-r^{i+1}$ are vertices of $\calQ-r^{i+1}$. Note that $r^i-r^{i+1}$ is zero in all but the last component, so this update takes only $t \leq k+1$ steps.

In order to warm-start the nearest point algorithm, the auxiliary matrix $R$ has to be recomputed as well. Using its definition
\[ R^TR = ee^T + V^T V\]
we can efficiently compute $R$ by Cholesky decomposition. After these updates we can directly start the nearest point algorithm in Step~2. Numerical experiments have shown that this speeds up LP decoding by a factor of two. In particular, the computation time of the Cholesky decomposition is negligible.

\section{The Complete Algorithm}

\prettyref{alg:mzfpNP} formalizes the procedure developed in \prettyref{sec:theory} in pseudocode.
\begin{algorithm}
\caption{Combinatorial Turbo LP Decoder (CTLP)}
\label{alg:mzfpNP}
\begin{algorithmic}[1]
\State Initialize edge cost $c(f)$ by \eqref{eq:edgeCost}
\State $f^0 \leftarrow \argmin\,\{c(f):\, f \in \Ppath\}$.
\State $v^0 \leftarrow \mathfrak D(f^0)$.
\State $ r^0 \leftarrow (0,\dotsc,0, v^0_{k+1})^T$\label{line:initialR}
\State $i \leftarrow 0$
\While{$v^i \neq r^i$}\label{line:npaMainLoop}
\State $v^{i+1} \leftarrow \op{NP}(r^i) = \argmin_{v \in \calQ} \norm{v - r^i}_2$
\label{line:nearestPoint}
\State $f^{i+1} = \mathfrak D^{-1}(v^{i+1})$
\label{line:preImage}
\State $a^{i+1} \leftarrow v^{i+1} - r^i$
\State $b^{i+1} \leftarrow a^{(i+1)T} v^{i+1}$
\State $r^{i+1} \leftarrow \left(0,\dotsc,0,
\frac{b^{i+1}}{a^{i+1}_{k+1}}\right)^T$
\State $i \leftarrow i + 1$
\EndWhile
\State \Return $f^i$
\end{algorithmic}
\end{algorithm}
The initial reference point $r^0$ is generated by first minimizing $c(f)$ on $\Ppath$ (thus solving TC-WS with $\gamma=(0,\dotsc,0,1)$ and projecting the result in constraints space onto $A_c$ (\prettyref{line:initialR}). Thereby we ensure that either $r^0 \notin \calQ$ or it is located on the boundary of $\calQ$, in which case it already is the optimal LP solution.
The solution of the nearest point problem and the recovery of the original flow are encapsulated in Lines~\ref{line:nearestPoint} and \ref{line:preImage}.

\section{Numerical Results}

\subsection{Running Time Comparison}
To evaluate the computational performance of our algorithm, we compare its running time with the commercial general purpose LP solver CPLEX \cite{CPLEX124} which is said to be one of the most competitive implementations available.

Simulations were run using LTE turbo codes with blocklengths 132, 228, and 396, respectively, and a three-dimensional turbo code with blocklength 384 (taken from \cite{Rosnes+Pseudo3DTC}) with various SNR values. For each SNR value, we have generated up to $10^5$ noisy frames, where the computation was stopped when $200$ decoding errors occured. This should ensure sufficient significance of the average results shown in Tables~\ref{table:lte40results}--\ref{table:3dresults}.
\begin{table}[ht]
	\caption{Average CPU time per instance (in $\frac1{100}$s of seconds) for the $(132,40)$ LTE turbo code}
	\label{table:lte40results}
	\centering
	\begin{tabular}{|c|c|c|c|c|c|c|}
		\hline
		SNR & $0$ & $1$ & $2$ & $3$ & $4$ & $5$\\
		\hline
		CPLEX & $9.1$ & $9.5$ & $9.6$ & $9.6$ & $9.6$ & $9.8$ \\
		CTLP & $1.4$ & $0.9$ & $0.5$ & $0.29$ & $0.24$ & $0.22$ \\
		\hline
		ratio & $6.5$  & $10.6$ & $19$ & $33$ & $40$ & $45$\\
		\hline
	\end{tabular}
\end{table}
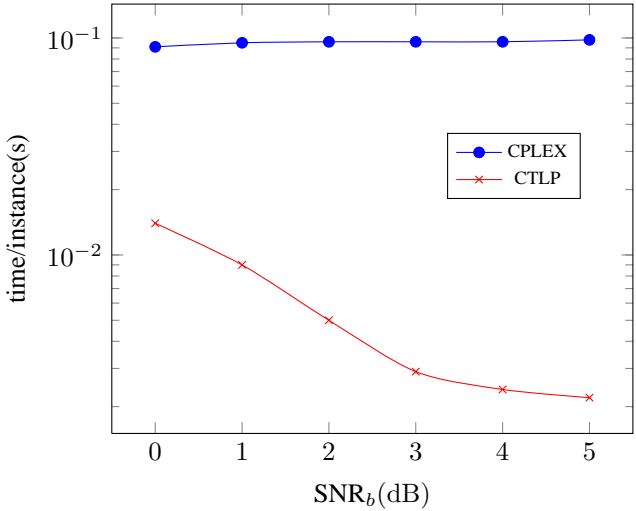
\begin{figure}
\begin{tikzpicture}

    \begin{semilogyaxis}[
        xlabel=$\text{SNR}_b(\op{dB})$,
        ylabel=time/instance(s),
        legend style={at={(.9,.7)},font=\scriptsize}]
    \addplot[smooth,mark=*,blue] plot coordinates {
        (0, 9.1e-2)
        (1, 9.5e-2)
        (2, 9.6e-2)
        (3, 9.6e-2)
        (4, 9.6e-2)
        (5, 9.8e-2)
    };
    \addlegendentry{CPLEX}

    \addplot[smooth,color=red,mark=x]
        plot coordinates {
            (0, 1.4e-2)
            (1, 0.9e-2)
            (2, 0.5e-2)
            (3, 0.29e-2)
            (4, 0.24e-2)
            (5, 0.22e-2)
        };
    \addlegendentry{CTLP}
    \end{semilogyaxis}

\end{tikzpicture}
\caption{CPU time comparison for the $(132,40)$ LTE Turbo code depending on the SNR value (note the logarithmic time scale).}
\end{figure}
\begin{table}[ht]
	\caption{Average CPU time per instance (in $\frac1{10}s$ of seconds) for the $(228,72)$ LTE turbo code}
	\label{table:lte72results}
	\centering
	\begin{tabular}{|c|c|c|c|c|c|c|}
		\hline
		SNR & $0$ & $1$ & $2$ & $3$ & $4$ & $5$\\
		\hline
		CPLEX($\times10^{-1}$) & $3.1$ & $3.4$ & $4.2$ & $4.6$ & $4.7$ & $4.7$ \\
		CTLP($\times10^{-1}$) & $0.7$ & $0.4$ & $0.15$ & $0.05$ & $0.04$ & $0.04$ \\
		\hline
		ratio & $4.4$  & $8.5$ & $28$ & $92$ & $118$ & $118$\\
		\hline
	\end{tabular}
\end{table}

\begin{table}[ht]
	\caption{Average CPU time per instance (in $\frac1{10}$s of seconds) for the $(396,128)$ LTE turbo code}
	\label{table:lte128results}
	\centering
	\begin{tabular}{|c|c|c|c|c|c|}
		\hline
		SNR & $0$ & $1$ & $2$ & $3$ & $4$\\
		\hline
		CPLEX& $4.4$ & $4.2$ & $3.6$ & $3.3$ & $3.2$\\
		CTLP & $6.3$ & $4.1$ & $0.6$ & $0.09$ & $0.08$\\
		\hline
		ratio & \textcolor{red}{$0.7$}  & $1$ & $6$ & $37$ & $40$\\
		\hline
	\end{tabular}
\end{table}

\begin{table}[ht]
	\caption{Average CPU time per instance (in seconds) for a $(384,128)$ 3-D turbo code}
	\label{table:3dresults}
	\centering
	\begin{tabular}{|c|c|c|c|c|c|}
		\hline
		SNR & $0$ & $1$ & $2$ & $3$ & $4$\\
		\hline
		CPLEX & $1.4$ & $1.2$ & $0.9$ & $0.72$ & $0.57$\\
		CTLP   & $4.5$ & $3.1$ & $0.8$ & $0.04$ & $0.014$\\
		\hline
		ratio & \textcolor{red}{$0.31$}  & \textcolor{red}{$0.39$} & \textcolor{red}{$1.1$} & $18$ & $41$\\
		\hline
	\end{tabular}	
\end{table}
As one can see, the benefit of using the new algorithm is larger for high SNR values. This becomes most eminent for the 3-D code for which the dimension of $\calQ$ is the highest, where the new algorithm is slower than CPLEX for SNRs up to $2$. The reason for this behavior can be explained by analyzing statistical information about various internal parameters of the algorithm when run with different SNR values:
\begin{itemize}
	\item The average dimension of the optimal nearest face, found in the last iteration of the algorithm, drops substantially with increasing SNR. Intuitively, it is not surprising that finding a face that needs less vertices to describe can be found more efficiently.
	\item In particular, the share of instances for which the LP solution is integral (and thus, the face dimension is zero) increases with the SNR.
	\item There are some trivial instances where the initial shortest path among both trellis graphs is already a valid codeword. This occurs more often for low channel noise and allows for extremely fast solution (no nearest point calculations have to be carried out).
	\item The average number of major cycles of the nearest point algorithm performed per instance is seen to drop rapidly with increasing SNR.
	\item Likewise, the the average number of main loops (\prettyref{line:npaMainLoop} of \prettyref{alg:mzfpNP}) drops, reducing the required calls to CTLP.
\end{itemize}
\prettyref{table:lte40stats} exemplarily contains the average per-instance values of these parameters for the $(132,40)$ LTE code and SNRs $0$, $2$, and $4$.
\begin{table}
	\caption{Statistical data for the $(132,40)$ LTE turbo code; average per-instance counts}
	\label{table:lte40stats}
	\centering
	\begin{tabular}{|c|c|c|c|c}
		\hline
		SNR        & $0$    & $2$    & $4$\\
		\hline
		face dim   & $25.2$ & $3.6$  & $0.01$\\
		integral   & $0.26$ & $0.89$ & $0.9995$\\
		trivial    & $0$    & $0.13$ & $0.64$ \\
		major      & $221$  & $53$   & $4$ \\
		main loops & $4.36$ & $1.9$ & $0.7$\\
		\hline
	\end{tabular}
\end{table}
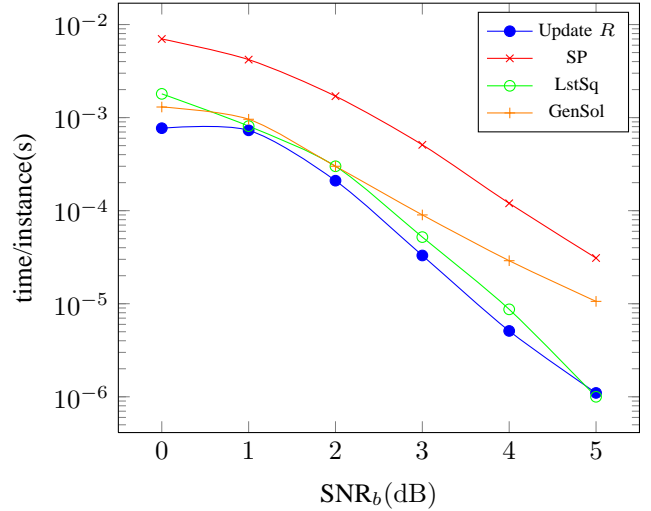
\begin{figure}
\begin{tikzpicture}

    \begin{semilogyaxis}[
        xlabel=$\text{SNR}_b(\op{dB})$,
        ylabel=time/instance(s),
        legend style={font=\scriptsize}]
    \addplot[smooth,mark=*,blue] plot coordinates {
        (0, 7.7e-4)
        (1, 7.3e-4)
        (2, 2.1e-4)
        (3, 3.3e-5)
        (4, 5.1e-6)
        (5, 1.1e-6)
    };
    \addlegendentry{Update $R$}

    \addplot[smooth,color=red,mark=x]
        plot coordinates {
            (0, 7e-3)
            (1, 4.2e-3)
            (2, 1.7e-3)
            (3, 5.1e-4)
            (4, 1.2e-4)
            (5, 3.1e-5)
        };
    \addlegendentry{SP}
    
    \addplot[smooth,color=green,mark=o]
        plot coordinates {
            (0, 1.8e-3)
            (1, 8.1e-4)
            (2, 3e-4)
            (3, 5.2e-5)
            (4, 8.7e-6)
            (5, 1e-6)
        };
    \addlegendentry{LstSq}
    
    \addplot[smooth,color=orange,mark=+]
        plot coordinates {
            (0, 1.3e-3)
            (1, 9.6e-4)
            (2, 3e-4)
            (3, 9e-5)
            (4, 2.9e-5)
            (5, 1.06e-5)
        };
    \addlegendentry{GenSol}
    
    \end{semilogyaxis}

\end{tikzpicture}
\caption{Average per-instance CPU time spent on various subroutines of the algorithm decoding the $(132,40)$ LTE code (SP=shortest path, LstSq=solution of least squares problems, GenSol=generation of solution in path space).}
\end{figure}

\subsection{Numerical Stability}
For larger codes, the dimension of $\calQ$ becomes very large which leads to numerical difficulties in the nearest point algorithm: the equation systems solved during the execution sometimes have rank “almost zero” which leads to division by very small numbers, resulting in the floating-point value \texttt{NaN}. Careful adjustment of the tolerance values for equivalence checks help to eliminate this problem at least for the block lengths presented in this numerical study.

In addition, it has proven beneficial to divide all objective values by $10$ in advance. Intuitively, this compresses $\calQ$ along the $c$-axis, evening out the extensiveness of the polytope in the different dimensions (note that for all axes other than $c$, the values only range from $-1$ to $1$).
\section{Improving Error-Correcting Performance}
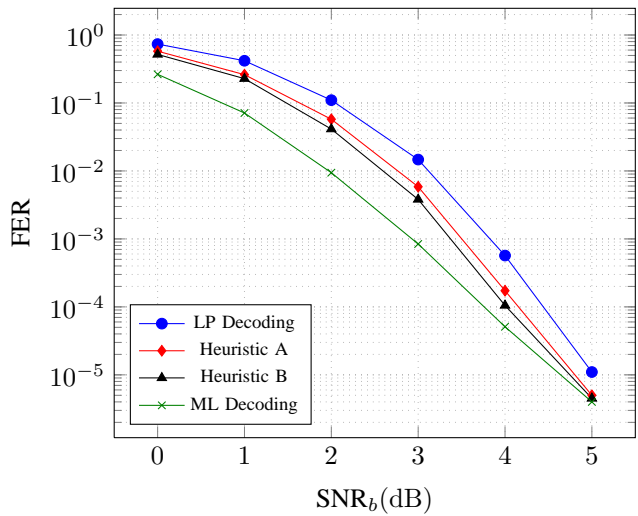
\begin{figure}
\pgfplotsset{grid style={help lines,dotted}}
\begin{tikzpicture}
  \begin{semilogyaxis}[
          xlabel=$\text{SNR}_b(\op{dB})$,
          ylabel=FER,
          grid=both,
          legend style={legend pos=south west,font=\scriptsize}]
      \addplot[mark=*,color=blue] plot coordinates {
          (0, 7.35e-1)
          (1, 4.18e-1)
          (2, 1.1e-1)
          (3, 1.47e-2)
          (4, 5.69e-4)
          (5, 1.1e-5)
      };
      \addlegendentry{LP Decoding}
      
      \addplot[mark=diamond*,color=red] plot coordinates {
          (0, 5.78e-1)
          (1, 2.61e-1)
          (2, 5.76e-2)
          (3, 5.85e-3)
          (4, 1.73e-4)
          (5, 5e-6)
      };
      \addlegendentry{Heuristic A}
      
      \addplot[mark=triangle*,color=black] plot coordinates {
          (0, 5.18e-1)
          (1, 2.28e-1)
          (2, 4.13e-2)
          (3, 3.81e-3)
          (4, 1.05e-4)
          (5, 4.5e-6)
      };
      \addlegendentry{Heuristic B}
      
      \addplot[mark=x,color=green!50!black] plot coordinates {
          (0, 2.64e-1)
          (1, 7.12e-2)
          (2, 9.42e-3)
          (3, 8.42e-4)
          (4, 5.10e-5)
          (5, 4e-6)
      };
      \addlegendentry{ML Decoding}
  \end{semilogyaxis}
\end{tikzpicture}
\caption{Decoding performance of the proposed heuristic enhancements on the ($132, 40$) LTE turbo code.}\label{fig:heuristics}
\end{figure}
As discussed above, \prettyref{alg:mzfpNP} can be easily modified to return a list of paths $f_i$, $i=1,\dotsc,t$, such that the LP solution is a convex combination of that paths. Each $f_i$ can be split into a paths $f_i^1$ and $f_i^2$ through trellis $T^1$ and $T^2$, respectively. A path in a trellis, in turn, can uniquely be extended to a codeword. Thus, we have a total of $2t$ candidate codewords. By selecting among them the codeword with minimum objective function value, we obtain a heuristic decoder (\emph{Heuristic A} in the following) that always outputs a valid codeword, and has the potential of a better error-correcting performance than pure LP decoding.

A slightly better decoding performance, at the cost of once more increased running time, is reached if we consider not only the paths that constitute the final LP solution but rather \emph{all} intermediate solutions of TC-WS. We call this modification \emph{Heuristic B}.

Simulation results for the ($132,40$) LTE code are shown in \prettyref{fig:heuristics}. As one can see, the frame error rate indeed drops notably when using the heuristics, but for low SNR values there still remains a substantial gap to the ML decoding curve. At $5\op{dB}$, Heuristic B empirically reaches ML performance; for lower SNR values it is comparable to a Log-MAP turbo decoder with 8 iterations.
\section{Conclusion and Outlook}
We have shown how the inherent combinatorial network-flow structure of turbo codes in form of the trellis graphs can be utilized to construct a highly efficient LP solver, specialized for that class of codes. The decrease in running time, compared to a general purpose solver, is dramatic, and in contrast to classical approaches based on Lagrangian dualization, the algorithm is guaranteed to terminate after a finite number of steps with the exact LP solution.

It is still an open question, however, if and how the LP can be solved in a \emph{completely} combinatorial manner. The nearest point algorithm suggested in this paper introduces a numerical component, which is necessary at this time but rather undesirable since it can lead to numerical problems in high dimension.

Another direction for further research is to examine the usefulness of our decoder as a building block of branch-and-bound methods that solve the integer programming problem, \ie, ML decoders. Several properties of the decoder suggest that this might be a valuable task. For instance, the shortest paths can be computed even faster if a portion of the varibales is fixed, or the algorithm could be terminated prematurely if the reference point exceeds a known upper bound at the current node of the branch-and-bound tree.

Finally, the concepts presented here might be of inner-mathematical interest as well. Optimization problems that are easy to solve in principle but have some complicating constraints are very common in mathematical optimization. Being able to efficiently solve their LP relaxation is a key component of virtually all fast exact or approximate solution algorithms.

\section*{Acknowledgements}
We would like to acknowledge the German Research Council (DFG), the German Academic Exchange Service (DAAD), and the Center for Mathematical and Computational Modelling ((CM)${}^2$) of the University of Kaiserslautern for financial support.

\bibliographystyle{IEEEtran}
\bibliography{IEEEabrv,abbr,agopt}
\end{document}